\documentclass[conference]{IEEEtran}
\usepackage{graphicx}
\usepackage{xcolor, tikz, pgfplots}
\usepackage{amsmath, amssymb}
\usepackage{caption}
\usepackage{subcaption}
\usepackage{verbatim}
\usepackage{float}

\newtheorem{theorem}{Theorem}
\newtheorem{prop}{Proposition}
\newtheorem{lemma}{Lemma}

\newcommand{\sibi}[1]{{\color{blue}#1}}
\usetikzlibrary{arrows}
\pgfplotsset{compat=1.17}

\title{On the Age of Information of a Queuing System with Heterogeneous Servers}

\author{\parbox{4 in}{\centering Anhad Bhati, Sibi~Raj~B~Pillai \\
\it Department of Electrical Engineering\\ Indian Institute of Technology Bombay, India\\
 {\tt\small \{anhadbhati,bsraj\}@ee.iitb.ac.in}}
   \parbox{3 in}{ \centering Rahul Vaze\\
 \it School of Technology and Computer Science\\
 Tata Institute of Fundamental Research, India\\
          {\tt\small rahul.vaze@gmail.com}}
 }

\IEEEoverridecommandlockouts
\IEEEpubid{\makebox[\columnwidth]{978-1-6654-4177-
3/21/\$31.00~\copyright{}2021 IEEE \hfill}
\hspace{\columnsep}\makebox[\columnwidth]{ }}

\begin{document}

\maketitle

\begin{abstract}

An optimal control problem with 
heterogeneous servers to minimize the average age of information (AoI) is considered. Each server maintains a separate queue, and each packet arriving to the system is randomly routed to one of the servers.  Assuming Poisson arrivals and exponentially distributed service times, we first derive an exact expression of the average AoI for two heterogeneous servers. Next, to solve for the optimal average AoI, a close approximation is derived, called the \emph{approximate AoI}, this is shown to be useful for multi-server systems as well. We show that for the optimal approximate AoI, server utilization (ratio of arrival rate and service rate) for each server should be same as the optimal server utilization with a single server queue. For two identical servers, it is shown that the average AoI is approximately $5/8$ times the average AoI of a single server. Furthermore, the average AoI is shown to decrease considerably with the addition of more servers to the system.
    %
\end{abstract}


\section{Introduction}
A classical problem in queuing systems is the optimal control with multiple 
heterogeneous servers to minimize the average delay (sojourn times) \cite{LinKumar}, where the decision variable chooses the server for serving the  head of the line packet.
In seminal work \cite{LinKumar}, it has been shown that with two servers performing non-preemptive service, the optimal policy is of a threshold type: the faster server never idles, and a packet is routed to the slower server only if the number of packets waiting in the queue is above a threshold. 
In addition to the usual performance metrics such as average delay and sojourn time, modern applications, e.g. IoT etc, require timely updates, and a metric which captures that neatly is known as the Age of Information (AoI). For a single server queue with arrivals,  AoI of several service disciplines like first come first serve (FCFS), last come first serve (LCFS) etc. have been analyzed in \cite{KaulYates}. A more recent comprehensive analysis can be found in \cite{InMaTa20}. 

\begin{figure}[htbp]
\centering
\includegraphics[width=\linewidth]{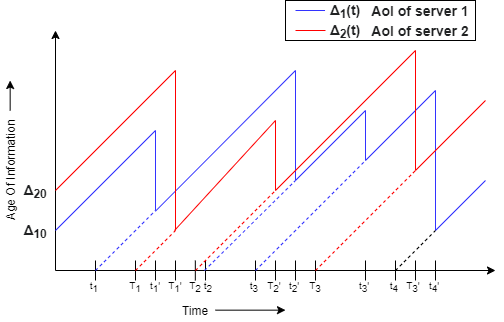}
\caption{AoI versus time: $t_i$ and $T_i$ ( $t_i'$ and $T_i'$) denote the respective arrival and departure instances of $i^{th}$ packet in the first (second) server.~\label{fig:aoi:graph}}
\end{figure}

In this paper, we are interested in computing the optimal average AoI of a multiserver system with 
heterogeneous servers. The threshold policy considered in \cite{LinKumar}  does not remain optimal when we change the performance metric from sojourn time to AoI. Therefore, we will consider a randomized scheduling policy instead.
Similar to \cite{LinKumar}, we consider that packets arrivals follow a Poisson distribution with rate $\lambda$, while the service distributions at the two servers are exponential with rates $\mu_1$ and $\mu_2$ respectively, where $\mu_1\ge \mu_2$, without loss of generality, with $\lambda < \mu_1+\mu_2$ for stability. For simplicity, we also enforce a non-preemptive FCFS disciple on each server. We wish to determine the optimal arrival rate to the multi-server system, as well as the optimal routing probability, which minimizes the average AoI.
When compared to the average delay case \cite{LinKumar}, the AoI metric behaves very differently compared to the average delay, thus bringing novel aspects to the problem. For example, in a M/M/1 queuing system, the AoI does not necessarily degrade with increasing arrival rate \cite{KaulYates}, as is the case with the average delay.
Moreover, with more than one server, the AoI can be shown to be the minimum  AoI among the individual servers (see Fig.~\ref{fig:aoi:graph} for a depiction), this is very different from average delay of multi-server systems.
Due to these differences, the classical result of \cite{LinKumar} does not apply directly for the considered problem. Whether a threshold type policy suffices for average AoI minimization is also unclear at this time. 
%
Thus, we instead focus our attention on  randomized routing policies, where each packet on its arrival is randomly routed to one of the servers.

\subsection{Prior Work}
To the best of our knowledge, there are only limited works on AoI minimization with multiple parallel servers, which we summarize as follows. 
A two server model, where the two servers are represented by a WiFi and cellular (4G) connection has been considered in
\cite{altman2019forever,el2012optimal} where one server is cheap but unreliable while the other is costly but reliable. The objective of the user is to minimize the average AoI of its messages taking into account its
utility and costs. An added feature over and above the model of \cite{altman2019forever,el2012optimal} where the two servers have different delays was studied in 
 \cite{Shroff3}. In \cite{altman2019forever,el2012optimal,Shroff3}, it is shown that the optimal policy is of threshold type. Another two server model has been studied in \cite{Clement}, where both servers have a common queue, with incoming packets being routed in FCFS fashion to the server which is available, and have provided an approximation for the average age, along with upper and lower bounds on the average AoI.
 There is of course a large body of work on AoI optimization with various models. For lack of space, however, we refer the reader to a recent survey \cite{yates2020age}, and the references there in.

\subsection{Our Contributions}
\begin{itemize}
    \item We derive an exact and closed form expression \eqref{eq:aoi:true} for the average AoI with two heterogeneous servers as a function of the randomized routing parameter $\alpha$. 
    \item Even though the exact expression for the average AoI is in closed form, it is not amenable to be optimized over the randomized routing parameter $\alpha$ and the arrrival rates. Thus, for analytical tractability, we first argue that the distribution of the AoI of an M/M/1 queue is well approximated by  a Gamma distribution with appropriate parameters. This leads to a pleasing formula for the approximate average AoI of two heterogeneous servers.
    \item We show that the optimal routing parameter $\alpha$ and the arrival rate $\lambda$ are such that $\rho_1= \frac{\alpha\lambda}{\mu_1} = \rho_2= \frac{(1-\alpha)\lambda}{\mu_2} = \rho^*$, where $\rho^*$ is the optimal server utilization for the M/M/1 queue \cite{KaulYates}. This result also naturally extends for minimizing approximate average AoI of more than two heterogeneous servers as well.
    \item One important take away from our analysis is that with two identical servers the average AoI is $5/8$ times the average AoI of a single server.  
\end{itemize}
%

\section{System Model} \label{sec:model}

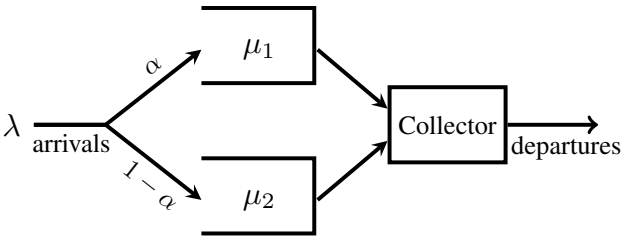
\begin{figure}[htbp]
\centering
\begin{tikzpicture}[line width=1.5pt]
\node (ar) at (-0.5,0) [left, scale=1.2]{$\lambda$};
\node (q1) at (2.5,1) [text width=1.0cm, text centered,scale=1.2]{$\mu_1$};
\node (q2) at (2.5,-1) [text width=1.0cm, text centered,scale=1.2]{$\mu_2$};
\node (cl) at (5,0) [rectangle, draw, minimum height=1cm]{Collector};
\draw[->,>=stealth] (ar) --++(1.25,0) -- node[above,sloped,pos=0.6]{$\alpha$}(q1.west);
\draw[->,>=stealth] (ar) --++(1.25,0) -- node[below,sloped, pos=0.6]{$1-\alpha$}(q2.west);
\draw (q1.north west) ++(0,0.25) --++(1.5,0) |- ++(-1.5,-1);
\draw (q2.north west) ++(0,0.25) --++(1.5,0) |- ++(-1.5,-1);
\draw[->,>=stealth] (q1.east) -- (cl.165);
\draw[->,>=stealth] (q2.east) -- (cl.195);
\draw[->] (cl) -- node[below, pos=0.65]{departures}++(2,0);
\node at (0,-0.25){arrivals};
\end{tikzpicture}

\caption{ Queuing System with two parallel servers ~\label{fig:sysmodel}}
\end{figure}

Consider the two server model shown in Fig.~\ref{fig:sysmodel}. We have two queues, namely $Q1$ and $Q2$, running in parallel. These two queues receive packets from a common Poisson arrival process with arrival rate $\lambda$. The packets arriving in the system are assigned to one of
the two queues, say with probabilities $\alpha$ and $(1-\alpha)$ respectively.
The servers are assumed to be non-identical, and service requirements are taken to be exponential. Let the first server have a service rate (we may call this also as speed of the server) of
$\mu_1$, whereas the second one has $\mu_2$.  Thus service times in server~$i$ are
exponential with average $\frac 1{\mu_i}$ for $i=1,2$.
W.l.o.g we take $\mu_1 \geq \mu_2$. Each server operates in a non-preemptive FCFS mode. Thus, a server should complete the service of all packets in the order that is assigned to it, with each job getting finished before taking up the next. The packets after completing their service are given to a common collector, which calculates the AoI based on the most recent packet available at that time instant, from either of the queues. It is possible that the packets will not complete their service in the same order that they arrived in the system, due to the presence of two parallel paths in the system.  The AoI status update is always computed causally with respect to the departed packet having the latest arrival time, irrespective of whether it came out of order or not. 

Our aim is to find out the average age of information of such a system, and find the value of the routing parameter $\alpha$ and the arrival rate $\lambda$ which will minimize the average AoI of this system.
In the $n$ server generalization of this problem, we have $n$ servers, and an arrival packet is assigned to server~$i$
with probability $\alpha_i$, $1 \leq i \leq n$. Again, we are interested in the minimum
average AoI.


Let us recollect  some of the existing single server results which are relevant to our analysis.
For the M/M/1 FCFS queue, Kaul et al in \cite{KaulYates} showed  that the average AoI with an arrival rate of $\lambda$ is given by
\begin{align} \label{eq:aoi:single}
\Delta = \frac{1}{\mu} \left(1 + \frac{1}{\rho} + \frac{\rho^2}{1-\rho}\right),
\end{align}
where $\mu$ and $\rho$ are the service rate and the server utilization respectively. Recall that $\rho= \frac\lambda\mu$.  Minimising this expression w.r.t. the utilisation shows that the average AoI is minimum  when
\begin{align}
    (1+\rho^2)(1-\rho)^2 = \rho^2.
\end{align}
Solving this we get the optimal server utilization as
\begin{align} \label{eq:rho:opt}
    \rho^* = \frac 12 \left( \sqrt{2} +1  - \sqrt{2\sqrt{2}-1 } \right),
\end{align}
which is about $53.17\%$. In reality, the distribution function of AoI carry more useful information than just the average value. Recently, the AoI distributions for various service disciplines were derived in \cite{InMaTa20}. For the M/M/1 FCFS queue,
the cumulative distribution function (CDF) of the $AoI$ is given by~\cite{InMaTa20}
\begin{multline} \label{eq:aoi:cdf}
 A(x) =1- e^{-\bar\rho \mu x}  +\left(\frac{1}{\bar \rho}+\rho \,\mu\, x\right) e^{-\mu x} - \frac{1}{\bar \rho}  e^{-\lambda x},
\end{multline}
where we took $\bar \rho = (1-\rho)$.

\section{Average AoI in  a Multi-Server   System} \label{sec:approx}
For simplicity, let us start with a two server model.
It is important to notice that the AoI of our two server system is calculated with respect to the combined departures. Thus, of the packets which have finished service, only the one with most recent arrival time can impact the AoI calculation. Equivalently, we can  calculate the individual AoIs separately with respect to the arrivals and departures at each server, and then declare the minimum of the two as the combined AoI. Thus the AoI of the system at instant $t$ is given by
$$\Delta (t) = \min\bigl(\Delta_{1}(t),\Delta_{2}(t)\bigr),$$
where $\Delta_i(t)$ is the AoI of queue~$i$ for $i=1,2$. Notice that the above result is independent of the arrival process.

Since the arrivals to the system happen according to a Poisson process of rate $\lambda$, the individual queues also observe independent Poisson arrivals, of rates $\lambda_1 =\alpha \lambda$
and $\lambda_2=(1-\alpha)\lambda$ respectively, due to the random thinning property of a Poisson process. Equivalently, $\frac{\lambda_i}{\lambda_1 + \lambda_2}$ is the probability that an arriving packet is assigned to queue~$i$, for $i=1,2$. Thus
\begin{align}
 P(\Delta(t)>x)  &= P(\min(\Delta_{1}(t),\Delta_{2}(t))>x) \notag \\ 
 &= P(\Delta_{1}(t)>x)P(\Delta_{2}(t)>x) .
\end{align}
Therefore, by knowing the distributions of $\Delta_{1}(t)$ and $\Delta_{2}(t)$, we can find $ P(\Delta(t)>x)$.
The average age of the whole system can now be written as  
\begin{align}\label{eq:general:exp}
\mathbb E \left[\Delta(t)\right] = \int_{0}^{\infty}  P\left(\Delta_{1}(t)>x\right)P\left(\Delta_2(t)> x\right)  \,dx.  
\end{align}
The ratio $\rho_i = \frac{\lambda_i}{\mu_i}, i=1,2$ has an important role in AoI calculations,  we refer to this quantity as the \emph{server loading} or \emph{utilization factor} of server~$i$.
Denoting $\bar \rho_i \triangleq 1 - \rho_i, i = 1,2$, and 
solving the above integral using the distribution in \eqref{eq:aoi:cdf}, we can compute the average AoI, which is stated
in the following propostion.
\def\bra{{\bar{\rho}}_1}
\def\brb{{\bar{\rho}}_2}

\begin{prop}
\begin{multline} \label{eq:aoi:true}
\mathbb E[ \Delta] = \frac{1}{\bra\mu_1 + \brb \mu_2} + \frac{1}{\bra\brb[\rho_1\mu_1 + \rho_2\mu_2]}+  
\frac{1}{\bra\brb(\mu_1 +\mu_2)} \\ + \frac{2\rho_1\rho_2\mu_1\mu_2}{(\mu_1 + \mu_2)^3 } +
 \frac 1{(\mu_1+\mu_2)^2}\left[\frac{\rho_2\mu_2}{\bra} + \frac{\rho_1\mu_1}{\brb} \right] -
 \Bigl\{ \frac{1}{\brb[\mu_2 + \bra\mu_1]} \\  + \frac{\rho_2\mu_2}{[\bra\mu_1 + \mu_2]^2} +
\frac{1}{\bra[\mu_1 + \brb\mu_2]} + \frac{\rho_1\mu_1}{[\brb\mu_2 + \mu_1]^2} \Bigr\}  \\ 
+
  \frac{1}{\brb[\rho_2\mu_2 + \bra\mu_1]} + 
 \frac{1}{\bra[\rho_1\mu_1 + \brb \mu_2]} 
+ \frac{\rho_1\mu_1}{\brb(\rho_2\mu_2 + \mu_1)^2} + \\ 
  \frac{\rho_2\mu_2}{\bra(\rho_1\mu_1 + \mu_2)^2}  
 -\frac{1}{\bra\brb}\left[\frac{1}{\mu_1 + \mu_2\rho_2}  + \frac{1}{\mu_2 + \mu_1\rho_1}\right] .
 \end{multline}
\end{prop}
\begin{IEEEproof}
An outline of the proof is provided in Appendix~\ref{sec:app:two}.
\end{IEEEproof}

While the AoI formula in \eqref{eq:aoi:true} is indeed in a closed form,
its rational expression  involves  high degree multinomials.
Thus, finding the optimal AoI for the given service rates appears less straightforward. Nevertheless, for the case where $\mu_1 = \mu_2$, we show that it is best to identically load each queue to a value slightly above the optimal utilization
of the individual queues.
\begin{lemma} \label{eq:sym:orig}
For $\mu_1= \mu_2$, the minimum average AoI is achieved at
$\rho_i = \rho^*, i=1,2$, where 
$\rho^*\approx 0.533391$.
\end{lemma}
\begin{IEEEproof}
The proof is included in Appendix~\ref{sec:app:sym}.
\end{IEEEproof}
Notice that the optimal utilization above is slightly larger than the $53.17\%$
for the single server M/M/1 queue. Thus, it is not optimal to choose the individual optima of the respective M/M/1 systems.
Unfortunately, such simplifications appear less evident when $\mu_1 \neq \mu_2$.
We now side-step the difficulty in handling the AoI formula in \eqref{eq:aoi:true} for non-identical servers, by providing a reasonably accurate and useful approximation to the AoI distribution for a single server.  
In particular, we approximate the individual AoI
distribution at each server by a Gamma distribution of parameter~$2$, such that average AoI at the server remains the same as before.


The implication of this approximation is as follows. Suppose the optimal pair for server utilization at the two servers is $(\rho_1^*, \rho_2^*)$, and the AoI distribution at each server at this operating point  is well approximated by the suggested gamma distribution. This will imply that the minimum average AoI of the approximated system cannot be more than a small amount from the actual minimum, as the approximation itself is expected to be close at the actual optimum. In other words, the approximation is expected to take us close to, or below, the true optimal value.

\begin{figure}
\centering
\begin{subfigure}[b]{0.24\textwidth}
\includegraphics[width=4.7cm, height=5cm]{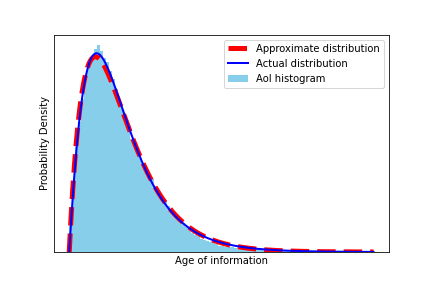}
\caption{$(\lambda,\mu) = (4,10)$}
\end{subfigure}
\begin{subfigure}[b]{0.24\textwidth}
\includegraphics[width=4.7cm, height=5cm]{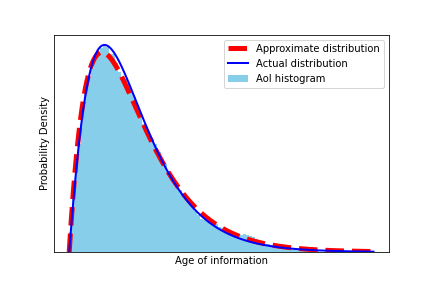}
\caption{ $(\lambda,\mu) = (5,10)$}
\end{subfigure}
\caption{AoI distribution and Gamma Approximation~\label{fig:gamma:demo}}
\end{figure}

In order to further convince the reader on the approximation, we show the actual and approximate AoI distributions for a single server M/M/1 system for two operating points,
in Fig.~\ref{fig:gamma:demo}. 
The blue curve represents the probability distribution function (PDF) corresponding to \eqref{eq:aoi:cdf}, along with its histogram from simulation as the shaded portion. The red dashed line shows the gamma approximation that we propose now.
Given the original single server queue with average AoI $2\Theta$, the respective PDF and  CDF of AoI can be approximated as:
\begin{align}\label{eq:gamma:pdf}
    f(x) &= \frac{x}{\Theta^2}\exp{\left(-\frac{x}{\Theta}\right)}, \,x \geq 0 \\
\label{eq:gamma:cdf}
    F(x) &= 1 - \exp{\left(-\frac{x}{\Theta}\right)}  - \frac{x}{\Theta}\exp{\left(-\frac{x}{\Theta}\right)},\, x \geq 0.
\end{align}
It is evident from Fig.~\ref{fig:gamma:demo} that the chosen gamma distribution closely approximates the actual AoI distribution for the
mentioned parameters. 
 In particular, this approximation turns out to be quite accurate when $\frac{\lambda}{\mu}$ 
values are not far away from $0.5$. However, we wish to remind the reader that the gamma approximation is not very effective for skewed queue utilization values ($\rho$ near $0$ or $1$), this can be inferred by comparing  \eqref{eq:aoi:cdf} and \eqref{eq:gamma:cdf}. 
Fortunately either too low or too high utilization seems less effective for AoI minimization in multi-server systems, for the same reasons as that of a single server queue~\cite{KaulYates}. 
While there may be other distributions which can be used as an approximation as well, the reason for choosing a gamma distribution in order to approximate our AoI distribution for a single server, as we will see going forward, is the simplistic expression for average AoI that it yields when we use this distribution to analyse the multi-server system.
The following lemma evaluates the average AoI under our approximation.
\begin{lemma}
For $i=1,2$, consider  two independent queues having AoI distributions according to \eqref{eq:gamma:cdf} with means $2\Theta_1$ and $2\Theta_2$ respectively. The average AoI $\hat {\mathbb E}[\Delta]$ with respect to the combined departures from the two queues is given by
\begin{align}\label{eq:AoI:approx}
\hat{\mathbb E}[\Delta]
    = 2 \Theta_0 \left( 1 + \frac{\Theta_0^2}{\Theta_1 \Theta_2}\right),
\end{align}
where
$$
\Theta_0 \triangleq \frac{\Theta_1 \Theta_2}{\Theta_1 + \Theta_2}.
$$
\end{lemma}
\begin{IEEEproof}
Using \eqref{eq:general:exp} and \eqref{eq:gamma:cdf}, the average AoI becomes
\begin{align} \label{eq:aoi:approx}
\hat{\mathbb E}[\Delta] 
&=\int_{0}^{\infty} (e^{-\frac{x}{\Theta_1}}  + \frac{x}{\Theta_1}e^{-\frac{x}{\Theta_1}})(e^{-\frac{x}{\Theta_2}}  + \frac{x}{\Theta_2}e^{-\frac{x}{\Theta_2}})  \,dx \notag \\
&=\int_{0}^{\infty} \left(e^{-\frac{x}{\Theta_0}}  + \frac{x}{\Theta_0}e^{-\frac{x}{\Theta_0}} + \frac{x^2}{\Theta_1\Theta_2}e^{-\frac{x}{\Theta_0}} \right)   \,dx \notag \\
& = \Theta_0  + \Theta_0 + 2 \frac{\Theta_0^3}{\Theta_1\Theta_2} .
\end{align}
\end{IEEEproof}

%
%

%
We term the average AoI function in \eqref{eq:AoI:approx} as the \emph{approximate} AoI of the original two server system. The advantage of the approximate value over the one given in \eqref{eq:aoi:true} is that the former is only a function of the individual average AoI at each of the queues. This makes the approximation much more convenient in computations, as  the following result exemplifies.

\begin{theorem}
For a given $\mu_1, \mu_2$, the approximate AoI 
$\hat{\mathbb E}[\Delta]$
is minimized at $(\rho_1, \rho_2) = (\rho^*, \rho^*)$, 
where $\rho^*$ is given in \eqref{eq:rho:opt}.
\end{theorem}
\begin{IEEEproof}
First, notice that
\begin{align} \label{eq:pd:1}
\frac{\partial \Theta_0}{\partial \rho_1} = 
\Theta_2 \frac{(\Theta_1+\Theta_2)\Theta_1^{\prime} - \Theta_1 \Theta_1^{\prime}}
{(\Theta_1 + \Theta_2)^2}.
\end{align}
Notice that in the interval $(0,1)$, $\Theta_1^{\prime} = 0$ only at $\rho_1=\rho^*$. Clearly,  the RHS of \eqref{eq:pd:1} is zero when $\rho_1= \rho^*$. Now, consider the partial derivative
\begin{align} \label{eq:pd:2}
\frac{\partial \hat{\mathbb E}[\Delta]}{\partial \rho_1}  = 2 \Theta_1^{\prime} +
    2\, \frac{2\Theta_0 \Theta_0^{\prime}(\Theta_1 + \Theta_2) - \Theta_1^{\prime}\Theta_0^2}
{(\Theta_1 + \Theta_2)^2}.
\end{align}
Clearly this is also zero at $\rho_1 = \rho^*$. Similar results hold for the partial derivative with respect to $\rho_2$
as well, completing the proof of the theorem.
\end{IEEEproof}

A notable advantage of our approximation is the elegant formula for the approximate average AoI in \eqref{eq:AoI:approx}. This form is much simpler in terms of analysis because it is in terms of the average AoI for the individual queues. In order to demonstrate its utility, let us find the optimal approximate AoI for the symmetric case with $\mu_1=\mu_2=\mu$.
Then the approximate average age is also a symmetric function with respect to  $\lambda_1$ and $\lambda_2$. We can minimize the average AoI under the additional constraint of $\lambda_1 + \lambda_2 = \lambda$, and and it is easy to see that the  minimum happens at the point $\lambda_1 = \lambda_2 = \frac{\lambda}2$. At this point, the respective average AoIs $2\Theta_1$ and $2\Theta_2$ for the two  queues are the same, and we take $\Theta_1 = \Theta_2 = \Theta$. which is half the average age of each queue. Using \eqref{eq:AoI:approx}, we get:
\begin{equation} \label{eq:sym:approx}
    \hat{\mathbb E}[\Delta]= \frac{5}{4} \,\Theta = \frac{5}{8} \bar{\Delta}_1,
\end{equation}
where $\bar\Delta_1$ is the average AoI of a single server M/M/1 queue.

Thus adding one more server in parallel can bring down the AoI to a factor of about $\frac{5}{8}$. Clearly  loading each queue with $\rho^* = \frac{\lambda}{2\mu}$ same
as that in \eqref{eq:rho:opt} is optimal in minimizing the approximate average AoI. Compare this against the value $\rho\approx 0.5333391$ for the actual system given in Lemma~\ref{eq:sym:orig}. 

For  queues with identical servers, calculating the true average AoI is more tedious than our approximation. However our approximation can easily handle several non-identical servers as well. To illustrate, let us extend our results to a system with $n$ heterogeneous servers. Let $\mu_i, 1 \leq i \leq n$ be the service rates, and $\alpha_i$ be the probability that an incoming packet is routed to server~$i$, where $\sum_{i = 1}^{i=n}\alpha_i = 1$. For a common Poisson arrival process of rate $\lambda=\sum_i \lambda_i$ to the multi-server system we have the following theorem.

\begin{theorem} \label{thm:multi}
For $n$ non-identical servers, the minimum approximate average AoI is achieved when $\frac{\alpha_i\lambda}{\mu_i} = \rho^*, 1 \leq i \leq n$, where $\rho^*$ is given in \eqref{eq:rho:opt}.
\end{theorem}
\begin{IEEEproof}
The proof is included in Appendix~\ref{sec:multi}.
\end{IEEEproof}
What we have shown is the optimal value of approximate AoI for a multi-server system happens when each server operates at its minimum AoI. Clearly, the common arrival rate and routing probability should facilitate the optimal operating condition for each server.
In short, the approximate AoI is minimized when each server utilization is slightly above $50\%$. We  also expect the true minimum average AoI to be not much different from that when the arrival rate to each of the servers is about half their respective service rates. 
Notice that when the utilization is close to 50\%, our approximation is a very valuable tool. On the other hand,  when the server utilization is lower, our approximation may not be that useful. Nevertheless, in the next section we show
that a small number of servers can considerably reduce the average AoI.

\section{Simulation Study} \label{sec:simul}

With the AoI distributions available, one can perform numerical comparisons to validate the AoI computations. Let us start with  a two server system and compare the approximate and actual AoIs.
Our first comparison is for the symmetric server case (Table I). The service rate here is kept $\mu = 20$ for both the queues, and the arrival rate to the system ($\lambda$) is varied from $0$ to $2\mu$, while assigning packets to the two queues with equal probabilities, i.e. both the queues have an effective arrival rate of $\frac \lambda 2$ each.
%
%
\begin{table}[]
\vspace*{0.35cm}
\begin{tiny}
\begin{center}
\resizebox{0.5\textwidth}{!}{
\begin{tabular}{| c | c | c | c |}
\hline
$\lambda$ & Empirical  & Approximate AoI & \% Error\\
&   average age &&\\
\hline
8 &	0.1691 & 0.1890	& 10.55\\
\hline
10 & 0.1467 & 0.1588 & 7.59\\
\hline
12 & 0.1347 & 0.1394 & 3.37\\
\hline
16 & 0.1197	& 0.1177 &	1.74\\
\hline
20 & 0.1102	& 0.1093 &	0.66\\
\hline
24 & 0.1094	& 0.1114 &	1.81\\
\hline
28 & 0.1271	& 0.1273 &	0.17\\
\hline
32 & 0.1572	& 0.1703 &	7.65\\
\hline
\end{tabular}
}
\end{center}
\caption{Empirical average v/s calculated Age, $\mu_1 = \mu_2 = 20$, $\lambda_1 = \lambda_2 = \lambda/2$}
\end{tiny}
\end{table}
\begin{table}[]
\begin{center}
\resizebox{0.49\textwidth}{!}{
\begin{tabular}{| c | c | c | c |}
\hline
$(\lambda_1,\lambda_2)$ & Actual AoI & Approximate AoI & \% Error\\
\hline
(20.735,0.465) &0.2312 & 0.2447 &   5.5057\\
\hline
(19.235,1.965) &0.1611 &   0.1730  &  6.8849\\
\hline
(17.735,3.465) &0.1323  &  0.1379  &  4.0641\\
\hline
(16.235,4.965) &0.1182  &  0.1196  &  1.1313\\
\hline
(14.735,6.465) &0.1114  &  0.1104  &  0.9948\\
\hline
(13.235,7.965) &0.1095  &  0.1075  &  1.8050\\
\hline
(11.735,9.465) &0.1117  &  0.1106  &  0.9697\\
\hline
(10.235,10.965) &0.1187  &  0.1206 &   1.5134\\
\hline
(8.735,12.465) &0.1335  &  0.1403  &  4.8964\\
\hline
(7.235,13.965) &0.1634  &  0.1745  &  6.3500\\
\hline
\end{tabular}
}
\end{center}
\caption{Actual Vs Approximate AoIs, $(\mu_1,\mu_2) = (25,15)$}
\end{table}
Observe that the approximate average AoI  appear close to  the actual average for values of $\lambda$ such that the utilisation of each queue is  around half, whereas some mismatch kicks in as the utilisation moves towards the extreme values.
The second comparison is for an asymmetric case, the first server having a service rate $\mu_1$ = 25 and the second server having a service rate of $\mu_2 = 15$ (Table II). The arrival rate is kept constant, $\lambda = 0.53(\mu_1+\mu_2)$, and the probability with which incoming packets are assigned to each server is varied, which is equivalent to varying the effective arrival rates of the individual servers while keeping the total arrival rate of the two server system to be constant.

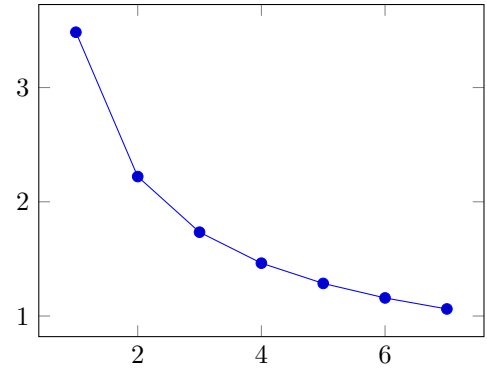
\begin{figure}[htbp]
\centering
\begin{tikzpicture}
\begin{axis} [height=6cm, width=7.5cm]
\addplot coordinates {
(1, 3.48452825904405)
(2, 2.22096515246663)
(3, 1.73391654867913)
(4, 1.46273790772633)
(5, 1.28542705303303)
(6, 1.15841723229711)
(7, 1.06191125473775)};
\end{axis}
\end{tikzpicture}
\caption{Average AoI Vs Number of Servers, $\mu_i=1, \forall i$. \label{fig:multi}}
\end{figure}

We have already seen that the addition of a server brings down the age to approximately $\frac 58$ of the original value. It is of interest to see how these results extend to more than two servers. Fig.~\ref{fig:multi} shows the average AoI reduction as we add more unit speed FCFS servers with random routing. Here the average AoI is computed with the utilization of each server kept around $0.53$.
The important thing to note here is that the multi-server average AoI quickly levels off with
the number of identical servers. Thus having a few servers itself may meet practical objectives on average AoI.

\section{Conclusion}\label{sec:conc} We have analyzed a multi-server queuing system with random routing from the point of view of average age of information. The benefits of having an additional server is shown, and it is interesting to note that the server utilization for minimal average AoI remains roughly the same as the single server M/M/1 queue. Future works will consider more general constraints on the weighted sum of arrival rates, as well as routing policies which take the queue states into account.

\bibliographystyle{unsrt}
\bibliography{refs.bib}
\begin{appendices}
\section{Calculation of AOI in \eqref{eq:aoi:true}}\label{sec:app:two}
Once the AoI distribution for a single server is known, We can use \eqref{eq:general:exp} to find the average AoI for the two server system. Let the AoI for server 1 and 2 be $\Delta_1(t)$ and $\Delta_2(t)$ respectively. Denote $(1-\rho_1)$ and $(1-\rho_2)$ as $\Bar{\rho_1}$ and $\Bar{\rho_2}$, respectively. From \eqref{eq:aoi:cdf},  for $i=1,2$,
\begin{equation}\label{eq:AoI:tail}
    P(\Delta_i(t)>x) = e^{-\bar{\rho_i}\mu_ix} - (\frac{1}{\bar{\rho_i}} + \rho_i\mu_ix)e^{-\mu_ix} + \frac{1}{\Bar{\rho_i}}e^{-\rho_i\mu_ix}.
\end{equation}
Let the AoI of the two server system be $\Delta(t)$. Then
\begin{align}\label{eq:twoserver:tail}
 P(\Delta(t)>x)  
 &= P(\Delta_{1}(t)>x)P(\Delta_{2}(t)>x) \notag\\
 &= e^{-(\Bar{\rho_1}\mu_1 + \Bar{\rho_2}\mu_2)} + \bigl( \frac{1}{\Bar{\rho_1}\Bar{\rho_2}} + \frac{\rho_2}{\Bar{\rho_1}}\mu_1x + \frac{\rho_1}{\Bar{\rho_2}}\mu_2x \notag\\
 &+ \rho_1\rho_2\mu_1\mu_2\bigr)e^{-(\mu_1+\mu_2)x} + G(\mu_1,\mu_2,\rho_1,\rho_2) \notag\\ 
 &+ G(\mu_2,\mu_1,\rho_2,\rho_1),
\end{align}
where,
\begin{align}
    G(\mu_1,\mu_2,\rho_1,\rho_2) = -(\frac{1}{\Bar{\rho_2}} + \rho_2\mu_2x)e^{-(\mu_2 + \Bar{\rho_1\mu_1})x} \notag \\
    + \frac{1}{\Bar{\rho_2}}e^{-(\bar{\rho_1}\mu_1 + \rho_2\mu_2)x} - \frac{1}{\Bar{\rho_2}}(\frac{1}{\Bar{\rho_1}} + \rho_1\mu_1x)e^{-(\mu_1 + \rho_2\mu_2)x}.
\end{align}
We can now compute the average AoI by
\begin{equation}
    \mathbb E \left[\Delta(t)\right] = \int_{0}^{\infty}  P\left(\Delta(t)>x\right)  \,dx.
\end{equation}
Evaluating this integral using \eqref{eq:twoserver:tail}  will
give the formula
in \eqref{eq:aoi:true}.

\section{Symmetric Servers} \label{sec:app:sym}


With $\mu_1 = \mu_2 = \mu$, the average AoI expression given in \eqref{eq:aoi:true} can be simplified considerably. In particular, the function $g(\rho_1, \rho_2) = \mu \mathbb E[\Delta]$ is easily seen to depend only on $\rho_1$ and $\rho_2$. Thus, we can take $\mu_1= \mu_2=1$, and find the optimal $(\rho_1, \rho_2)$.

In order to proceed with the optimization, we first impose an additional constraint that $\rho_1+\rho_2 = 2\rho$, and later relax this constraint to allow for all values of $\rho <  \mu$.  Notice that the function $g(\rho_1, \rho_2)$ is symmetric in the arguments, and thus its derivative along the line $\rho_1+\rho_2 \leq 2 \rho$ will vanish when $\rho_1= \rho_2 = \rho$. With $\rho_1=\rho_2=x$, let the function $f(x) = g(x, x)$ represent the  average AoI in \eqref{eq:aoi:true}. The derivative of $f(x)$
is the rational function  given in \eqref{eq:der:1},  at the bottom of this page.

\begin{center}
\begin{tikzpicture}[remember picture, overlay]
\node at (current page.south)[text width=16.5cm, yshift=6.0cm, rectangle, draw]{
\begin{align} \label{eq:der:1}
f^{\prime}(x) = \frac{x^{11} - 5 x^{10} + 4 x^{9} + 24 x^{8} - 27 x^{7} + 30 x^{6} - 123 x^{5} + 169 x^{4} - 71 x^{3} - 14 x^{2} - 4 x + 8}{2 x^{2} \left(x - 2 \right)^{3}  \left(x - 1 \right)^{2} \left(x + 1\right)^{3}}.
\end{align}
};
\end{tikzpicture}
\end{center}
The only zero of the function $f^{\prime}(x)$ in the interval $(0,1)$ happens at $x\approx 0.533391$, whereas the second derivative $f^{\prime\prime}(x)$ has no zeros in $(0,1)$. 
This effectively proves the lemma.

\newpage
\section{Multi-server System in Theorem~\ref{thm:multi}} \label{sec:multi}

Now let us consider an extension of our two server systems to $n$ servers running in parallel. By our random routing, each server observes an independent Poisson process of rate $\lambda_j = \alpha_j \lambda$, where $\alpha_j$ is the routing probability to server~$j$, and $\lambda$ is the combined arrival rate. 
%
For $j = 1,2,...,n$, let $\Delta_j(t) $ and $2\Theta_j$ be the instantaneous and average age of the $j^{th}$ server. For the $j^{th}$ server, we take the tail distribution of AoI according to the gamma approximation, i.e.,
\begin{equation} \label{eq:app:prob}
    P(\Delta_j(t) > x) = \exp{\left(-\frac{x}{\Theta_j}\right)}  + \frac{x}{\Theta_j}\exp{\left(-\frac{x}{\Theta_j}\right)},\, x \geq 0.
\end{equation}


For convenience, let us denote the RHS of \eqref{eq:app:prob} as $G(x,\Theta_j)$. Taking derivatives
\begin{align}
    \frac{d G(x,\Theta)}{d \Theta} &= \frac{x}{\Theta^2}e^{-\frac{x}{\Theta}}(1+\frac{x}{\Theta}) - \frac{x}{\Theta^2}e^{-\frac{x}{\Theta}} \notag\\
    &= \frac{x^2}{\Theta^3}e^{-\frac{x}{\Theta}} \\
    &\geq 0
\end{align}
Therefore, if $\Theta_1>\Theta_2$, then $G(x,\Theta_1) > G(x,\Theta_2)$. 
%

The approximate average AoI
can now be computed as
\begin{align}\label{eq:age:ineq}
\hat{\mathbb E}[\Delta] &= \int_0^{\infty} P(\Delta(t) > x) dx \\
    &= \int_0^{\infty}\prod_{j = 1}^{n}P(\Delta_j(t)>x)\notag\\
    &=  \int_0^{\infty} \prod_{j = 1}^{n}G(x,\Theta_j) dx.
\end{align}
Since $G(x,\Theta)$ is non-negative and  monotonously non-increasing in $\Theta$,  the above RHS is clearly minimized by choosing the minimal value for each $\Theta_j$, $1 \leq j \leq n$.  This proves the theorem.

\end{appendices}

\end{document}